\documentclass[pra,aps,amsmath,amssymb,amsfonts,twocolumn,nofootinbib]{revtex4}
\usepackage{amssymb}
\usepackage{}
\usepackage{bm,mathrsfs}

\usepackage{graphicx}
\usepackage{epsfig}
\usepackage{amsmath,bbm}
\usepackage{amsfonts,amssymb}
\usepackage{times}
\usepackage{verbatim}
\usepackage[sort&compress]{natbib}

\usepackage{amsmath}
\usepackage[colorlinks,breaklinks,linkcolor=blue,anchorcolor=blue,citecolor=blue,urlcolor=blue,
dvipdfm
]{hyperref}

\newcommand{\be}{\begin{equation}}
\newcommand{\ee}{\end{equation}}
\newcommand{\bea}{\begin{eqnarray}}
\newcommand{\eea}{\end{eqnarray}}

\def\>{\rangle}
\def\<{\langle}

\def\qed{\leavevmode\unskip\penalty9999 \hbox{}\nobreak\hfill
     \quad\hbox{\leavevmode  \hbox to.77778em{%
               \hfil\vrule   \vbox to.675em%
               {\hrule width.6em\vfil\hrule}\vrule\hfil}}
     \par\vskip3pt}
     
\begin{document}

\newtheorem{theorem}{Theorem}
\newtheorem{lemma}[theorem]{Lemma}
\newtheorem{corollary}[theorem]{Corollary}
\newtheorem{proposition}[theorem]{Proposition}
\newtheorem{definition}[theorem]{Definition}
\newtheorem{example}[theorem]{Example}
\newtheorem{conjecture}[theorem]{Conjecture}

\title{Dissipation induced $W$ state in a Rydberg-atom-cavity system}

\author{Dong-Xiao Li}
\affiliation{Center for Quantum Sciences and School of Physics, Northeast Normal University, Changchun, 130024, People's Republic of China}
\affiliation{Center for Advanced Optoelectronic Functional Materials Research, and Key Laboratory for UV Light-Emitting Materials and Technology
of Ministry of Education, Northeast Normal University, Changchun 130024, China}

\author{Xiao-Qiang Shao\footnote{Corresponding author: shaoxq644@nenu.edu.cn}}
\affiliation{Center for Quantum Sciences and School of Physics, Northeast Normal University, Changchun, 130024, People's Republic of China}
\affiliation{Center for Advanced Optoelectronic Functional Materials Research, and Key Laboratory for UV Light-Emitting Materials and Technology
of Ministry of Education, Northeast Normal University, Changchun 130024, China}

\author{Jin-Hui Wu}
\affiliation{Center for Quantum Sciences and School of Physics, Northeast Normal University, Changchun, 130024, People's Republic of China}
\affiliation{Center for Advanced Optoelectronic Functional Materials Research, and Key Laboratory for UV Light-Emitting Materials and Technology
of Ministry of Education, Northeast Normal University, Changchun 130024, China}

\author{X. X. Yi}
\affiliation{Center for Quantum Sciences and School of Physics, Northeast Normal University, Changchun, 130024, People's Republic of China}
\affiliation{Center for Advanced Optoelectronic Functional Materials Research, and Key Laboratory for UV Light-Emitting Materials and Technology
of Ministry of Education, Northeast Normal University, Changchun 130024, China}
\date{\today}

\begin{abstract}
A dissipative scheme is proposed to prepare tripartite $W$ state in a Rydberg-atom-cavity system. It is an organic combination of quantum Zeno dynamics, Rydberg antiblockade and atomic spontaneous emission to turn the tripartite $W$ state into the unique steady state of the whole system. The robustness against the loss of cavity and the feasibility of the scheme are demonstrated thoroughly by the current experimental parameters, which leads to a high fidelity above $98\%$.
\end{abstract}

\maketitle

With the rapid development of quantum information, different specific tasks appear and require various multipartite entanglements \cite{pra062315ref8,pra062315ref14}. An crucial representative of multipartite entanglements is the $W$ state \cite{PhysRevA.62.062314}, which exhibits the high robustness against the loss of particle: While all parts but two parts of the $N$-part $W$ state are collapsed by the measurement, the state of the remaining two-qubit system is still entangled. Therefore, the $W$ state is widely used in quantum information processing, such as quantum cloning machines \cite{chen2012ref20} and quantum memories \cite{pra042113ref22}, and multifarious schemes are put forward to prepare the $W$ state \cite{PhysRevA.83.050303,PhysRevA.87.013842,PhysRevLett.117.140502}. Fujii \textit{et al.} used linear optics and postselections to generate an $N$-qubit $W$ state among separated quantum nodes  \cite{PhysRevA.83.050303} and { Lin \textit{et al.} prepared the $W$ state with three trapped atomic ions by laser fields \cite{PhysRevLett.117.140502}.}

One of the primary obstacles of the generation of multipartite entanglements is the quantum noise from environment, which is inevitable and will result in the decoherence and dissipation \cite{pra042323ref4}. Nevertheless, the dissipative schemes of preparing entanglements improve the character of the quantum noise by considering it  as an important resource to achieve the goals, and now are in widespread uses \cite{fiberref38,chen2012,PhysRevA.87.042323,Shen:14,prl040501,Li:17}, \textit{e.g.} Kastoryano \textit{et al.} exploited the decay of cavity to prepare the maximally entangled state of two atoms in an cavity \cite{fiberref38}. Chen \textit{et al.} \cite{chen2012} and Sweke \textit{et al.} \cite{PhysRevA.87.042323} extended the method to dissipatively prepare $W$ state within context of cavity quantum electrodynamics. However, they both required precisely tailored condition to achieve the proposal. In addition, Reiter \textit{et al.} dissipatively produced many-body entanglements of atoms coupled to common oscillator modes \cite{prl040501}, which required many classical and quantized fields.

The Rydberg blockade effect has recently become of interest in the literature \cite{prl013001ref11}. The phenomena of Rydberg blockade effect can be intuitively understood by a blockade sphere \cite{pra033422ref1}. Within the radius of blockade sphere, the dipole-dipole interaction will induce a large energy shift to significantly suppress two or more Rydberg atoms excited simultaneously. On the other hand, an opposite effect, the Rydberg antiblockade, has also been predicted by Ates \textit{et al.} in a three-level two-photon Rydberg excitation scheme, which induces the simultaneous excitations of two Rydberg atoms and the inhibition of Rydberg blockade \cite{klmoeref30}. The antiblockade effect provides a new perspective for quantum entanglement and multiqubit logic gates \cite{klmoeref32,klmoeref35,pra062315}. Quite recently, Shao \textit{et al.} designed a simplified proposal through Rydberg atoms to prepare a Greenberger-Horne-Zeilinger (GHZ) state by dissipation \cite{pra062315} with fewer classical and quantized fields than those in \cite{prl040501}.

\begin{figure}[h]
\centering
\includegraphics[scale=0.13]{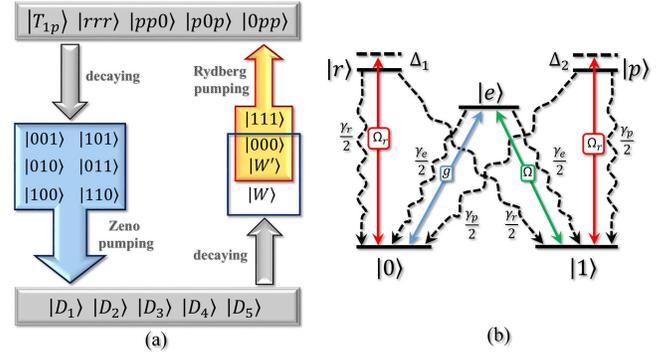}
\caption{\label{flowgraph}(a) The flow chart of the scheme, where $|W'\rangle=(|110\rangle+|101\rangle+|011\rangle)/\sqrt{3}, |D_1\rangle=(|00e\rangle-|e00\rangle)/\sqrt{2},
|D_2\rangle=(2|0e0\rangle-|e00\rangle-|00e\rangle)/\sqrt{6},
|D_3\rangle=(|10e\rangle-|1e0\rangle)/\sqrt{2},
|D_4\rangle=(|01e\rangle-|e10\rangle)/\sqrt{2},
|D_5\rangle=(|e01\rangle-|0e1\rangle)/\sqrt{2},$ and $|T_{1p}\rangle=(|1pp\rangle+|p1p\rangle+|pp1\rangle)/\sqrt{3}$. (b) Atomic level configuration.}
\end{figure}
In this work, we consider a dissipative scheme to realize a tripartite $W$ state, $|W\rangle=(|100\rangle+|010\rangle+|001\rangle)/\sqrt{3}$, with an optical cavity trapping three five-level Rydberg atoms which include two ground states $|0\rangle$ and $|1\rangle$, an excited state $|e\rangle$, and two Rydberg states $|r\rangle$ and $|p\rangle$.
The flow chart of the present scheme is plotted in Fig.~\ref{flowgraph}(a). It can be separated into two simultaneous processes: (i) The Zeno pumping aims to pump the states with one or two atoms in $|0\rangle$ into the excited states $|D_1\rangle,|D_2\rangle,|D_3\rangle,|D_4\rangle$ and $|D_5\rangle$, which further decay to $\{ |000\rangle,~|W'\rangle,~|W\rangle \}$ by the atomic spontaneous emission. (ii) The purpose of the Rydberg pumping is to drive the states $\{ |000\rangle,~|W'\rangle,~|111\rangle \}$ to the excited states $|T_{1p}\rangle,|rrr\rangle,|pp0\rangle,|p0p\rangle$ and $|0pp\rangle$, which can also spontaneously emit to the states with one or two atoms in $|0\rangle$. The two simultaneous processes create a cycle between all ground states except $|W\rangle$ and lead to the system stable at $|W\rangle$ finally. On the whole, the present scheme has fourfold features: (i) The spontaneous emission is a powerful resource to achieve target state and the decay of cavity is availably depressed by quantum Zeno dynamics in Zeno pumping. (ii) The realization of the scheme is independent of initial state. { (iii) Compared with \cite{chen2012}, \cite{prl040501} and \cite{pra062315}, it is easier to operate in experiment.   There needs no precisely tailored Rabi frequencies, coupling strength between cavity and atoms, or atom-dependent light shift of ground state in our proposal.} And the $W$ state can be accomplished with fewer classical and quantized fields. (iv) The fidelity of the $W$ state can be above $98\%$ with the current experimental parameters.

\begin{figure}
\centering
\includegraphics[scale=0.13]{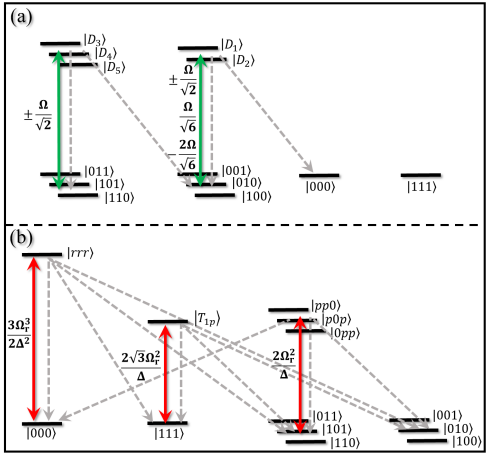}
\caption{\label{effzeno} (a) The effective transitions of Zeno pumping. (b) The effective transitions of Rydberg pumping.}
\end{figure}
The corresponding atomic level is shown in Fig.~\ref{flowgraph}(b) and it can be divided into two parts: (i) The ground states $|0\rangle$ and $|1\rangle$ are coupled resonantly to the excited state $|e\rangle$ by a quantized cavity mode with strength $g$ and a classical laser field with Rabi frequency $\Omega$, respectively. For simplicity, we consider the excited states decay to the ground states with the same branching ratio $\gamma_e/2$. (ii) The transitions $|0\rangle\leftrightarrow|r\rangle$ and $|1\rangle\leftrightarrow|p\rangle$ are dispersively driven by two independent classical fields with the same Rabi frequency $\Omega_r$ and different detuning $\Delta_1=-2\Delta,~\Delta_2=-\Delta$, respectively. The Rydberg state $|r(p)\rangle$ decays to the ground states with the same rate $\gamma_{r(p)}/2$ and in what follows, we consider $\gamma_r=\gamma_p=\gamma$. In the interaction picture, the corresponding Hamiltonian $H_I$ can be written as ($\hbar=1$) $
H_I=H_I^Z+H_I^R,$ with
$H_I^Z=\sum_{i=1}^3\Omega|e\rangle_i\langle1|+g|e\rangle_i\langle0|a+{\rm H.c.},$ and $H_I^R=\sum_{i=1}^3\Omega_r(|r\rangle_i\langle0|e^{-2i\Delta t}+|p\rangle_i\langle1|e^{-i\Delta t}+{\rm H.c.})+\sum_{i<j}U_{rr}(|rr\rangle_{ij}\langle rr|+|pp\rangle_{ij}\langle pp|)+U_{rp}(|pr\rangle_{ij}\langle pr|+|rp\rangle_{ij}\langle rp|),$
where $a$ denotes the annihilation operator of the quantized cavity mode, and $H_I^Z$ and $H_I^R$ represent the Zeno pumping and the Rydberg pumping, respectively. The Rydberg-mediated interactions are described by $U_{pp}=U_{rr}$ when both atoms occupy the same Rydberg states and $U_{pr}=U_{rp}$ when the two atoms occupy the different Rydberg states, and we consider the interaction $U^{ij}$ of $i$-th and $j$-th atoms as $U_{rr}^{(12)}=U_{rr}^{(23)}=U_{rr}^{(31)}=U_{rr}$.  Generally speaking, the interaction energy lies on angular degrees of freedom, which leads to the distinction of the coupling strengths when $|r\rangle,~|p\rangle$ represent different Zeeman sublevels, \textit{i.e.} $U_{rr}\neq U_{rp}$ \cite{prl033607ref15}.  When we choose the suitable principle quantum number  \cite{prl033607ref15,pra043802}, the Rydberg-mediated interactions satisfy $U_{rr}\gg U_{rp}$  and the last term of $H_I^R$ can be neglected. Then the Hamiltonian reads as
$
H_I=H_I^Z+H_I^R,
$
where $H_I^Z=\sum_{i=1}^3\Omega|e\rangle_i\langle1|+g|e\rangle_i\langle0|a+{\rm H.c.},H_I^R=\sum_{i=1}^3$
$\Omega_r(|r\rangle_i\langle0|+|p\rangle_i\langle1|+{\rm H.c.})-2\Delta|r\rangle_i\langle r|-\Delta|p\rangle_i\langle p|+\sum_{i<j}U_{rr}(|rr\rangle_{ij}\langle rr|+|pp\rangle_{ij}\langle pp|).$
The decay of $i$-th atom and cavity can be described by Lindblad operators as
$L_i^{1(2)}=\sqrt{\gamma/2}|0(1) \rangle_i\langle r|,L_i^{3(4)}=\sqrt{\gamma/2}|0(1)\rangle_i\langle p|,$
$L_i^{5(6)}=$
$\sqrt{\gamma_e/2}|0(1)\rangle_i\langle e|,$ $L^{c}=\sqrt{\kappa}a.$
Combining the Hamiltonian of system and the Lindblad operators, the Markovian master equation of system can be obtained as
$
\dot\rho=$
$-i[H_I,\rho]+\mathcal{L}^a\rho+\mathcal{L}^c\rho,
$
where the superoperators, $\mathcal{L}^a\rho=$
$\sum_{k=1}^6\sum_{i=1}^3L_i^k\rho L_i^{k\dag}-(L_i^{k\dag} L_i^k\rho+\rho L_i^{k\dag} L_i^k)/2$ and $\mathcal{L}^c\rho=$
$L^c\rho L^{c\dag}-(L^{c\dag} L^c\rho+\rho L^{c\dag} L^c)/2$, denote the decay of the atoms and cavity, respectively.

In order to interpret the physical principle of the present scheme, we analyze the effective Hamiltonian of the Zeno pumping and the Rydberg pumping in detail. The Hamiltonian of Zeno pumping, $H_I^Z$, can be reformulated as $H_I^Z=\Omega H^Z_1+gH^Z_2,$ where $H_1^Z=\sum_{i=1}^3|e\rangle_i\langle1|+{\rm H.c.}$ and $H_2^Z=\sum_{i=1}^3|e\rangle_i\langle0|a+{\rm H.c.}$. According to the rigorous extensions of the quantum Zeno dynamics \cite{myolref20,myolref22}, in the limit of $g\gg\Omega$, $H_I^Z$ can be further simplified as $H_I^Z=\sum_ngE_nP_n+\Omega P_nH_1^ZP_n$, where $P_n=|E_n\rangle\langle E_n|$ and $|E_n\rangle$ is  the eigenstate of $H_2^Z$ corresponding to the eigenvalue $E_n$.

{ If the system is initialized in the ground state with the cavity  in the vacuum state  $|0\rangle_c$, the system will only evolve in the subspace of $E_0=0$ corresponding to the basis $\{ |001\rangle,|010\rangle,|100\rangle,|011\rangle,|101\rangle,|110\rangle,|D_l\rangle  \}\otimes|0\rangle_c$, where $|D_l\rangle\otimes|0\rangle_c~(l=1,2,3,4,5)$ are the dark states of $H_2^Z$.} Then the Hamiltonian of Zeno pumping can be simplified as
$
H_{\rm eff}^Z=\Omega P_0H_1^ZP_0=\left(H_{\rm eff1}^Z+H_{\rm eff2}^Z\right)\otimes|0\rangle_c\langle0|,
$
where $H_{\rm eff1}^Z$
$=\Omega|001\rangle( \langle D_1|/\sqrt{2}-\langle D_2|/\sqrt{6}  )-\Omega|100\rangle( \langle D_1|/\sqrt{2}+$
$\langle D_2|/\sqrt{6}  )+2\Omega|010\rangle\langle D_2|/\sqrt{6}+{\rm H.c.}$ and $H_{\rm eff2}^Z=\Omega|101\rangle( \langle D_3|+$
$\langle D_5|  )/\sqrt{2}+\Omega|011\rangle( \langle D_4|-\langle D_5|  )/\sqrt{2}-\Omega|110\rangle( \langle D_3|+\langle D_4|$
$  )/\sqrt{2}+{\rm H.c.}.$ { The above results are similar to the effective Hamiltonian of simplified $Z$ pumping in \cite{pra062315}.  However, our Zeno pumping aims to stabilize the system to the subspace in the basis of $\{ |000\rangle,|111\rangle,|W'\rangle,|W\rangle  \}$, thus the atom-dependent light shift $\delta_i$ is unwanted.}
Based on the $H_{\rm eff}^Z$, the status of cavity has been restrained in the vacuum state resulting in the significant inhibition of the decay of cavity and we omit the effective Lindblad operator of cavity. Then we expand the atomic Lindblad operators, $L_i^k$, by $\{ |001\rangle,|010\rangle,|100\rangle,|011\rangle,|101\rangle,|110\rangle,|D_l\rangle  \}$, and obtain the effective Lindblad operators of atoms as
$L_{\rm eff}^{1(2)}=\sqrt{\gamma_e/12}$$|100(001)\rangle\langle D_1|,$$L_{\rm eff}^{3}$$=$$\sqrt{\gamma_e/3}$$|010\rangle\langle D_1|,$$L_{\rm eff}^{4(5)}$$=$$\sqrt{\gamma_e/2}$
$|000\rangle\langle D_{1(2)}|,$$L_{\rm eff}^{6(7)}$$=$$\sqrt{\gamma_e/4}$$|100(001)\rangle\langle D_2|,$$L_{\rm eff}^{8(9)}$$=$$\sqrt{\gamma_e/4}$
$|110(101)\rangle\langle D_3|,$$L_{\rm eff}^{10(11)}$$=$$\sqrt{\gamma_e/2}$$|100(010)\rangle\langle D_{3(4)}|,$$L_{\rm eff}^{12(13)}$$=$
$\sqrt{\gamma_e/4}$$|110(011)\rangle\langle D_4|,$$L_{\rm eff}^{14(15)}$$=$$\sqrt{\gamma_e/4}$$|101(011)\rangle\langle D_5|$ and $L_{\rm eff}^{16}=\sqrt{\gamma_e/2}|001\rangle\langle D_5|.$ The effective Markovian master equation of the Zeno pumping can be written as $\dot\rho=-i[H_{\rm eff}^Z,\rho]+\mathcal{L}_{\rm eff}\rho,$ where $\mathcal{L}_{\rm eff}\rho=\sum_{k=1}^{16}L_{\rm eff}^k\rho L_{\rm eff}^{k\dag}-\frac{1}{2}(L_{\rm eff}^{k\dag}L_{\rm eff}^k\rho+\rho L_{\rm eff}^{k\dag}L_{\rm eff}^k).$

In Fig.~\ref{effzeno}(a), we plot the effective transitions of Zeno pumping, where we have omitted the symbol of the cavity in vacuum state. As an example, we choose $|011\rangle$ as an initial state and investigate the corresponding evolution. Firstly, it will be pumped to the excited states $|D_4\rangle$ and $|D_5\rangle$ by the effective classical fields with Rabi frequencies $\pm\Omega/\sqrt{2}$, respectively. Then the excited state $|D_{4(5)}\rangle$ will decay to the ground states of one atom in $|0\rangle$ and two atoms in $|0\rangle$. On the one hand, the ground states of one atom in $|0\rangle$ will circulate the above processes. On the other hand, the ground states of two atoms in $|0\rangle$ will be driven to the excited states $|D_1\rangle$ and $|D_2\rangle$, which can revert to the ground states of two atoms in $|0\rangle$ or $|000\rangle$ by spontaneous emission. Finally, the system initialized in $|011\rangle$ will be stable at the steady subspace spanned by $|000\rangle,|W'\rangle$ and $|W\rangle$. As for the system initialized in $|111\rangle$, it will not evolve to any other state due to the limit of $g\gg\Omega$. In general, the system will eventually reach the mixed states of $|000\rangle,|111\rangle,|W'\rangle$ and $|W\rangle$ with an arbitrary initial state.

To realize the preparation of $|W\rangle$, we also need to utilize the Rydberg pumping to drive the states $|000\rangle,|111\rangle$ and $|W'\rangle$ into excited states and the target state $|W\rangle$ becomes the unique steady state of total system. We select the strength of Rydberg interaction $U_{rr}=2\Delta$ and $\Delta\gg\Omega_r$ to realize the Rydberg antiblockade. Neglecting the order of $\mathcal{O}(\Omega_r^2/\Delta^2)$, the total effective Hamiltonian of Rydberg pumping can be obtained as
$
H_{\rm eff}^R=H^{R_0}_{\rm eff}+H^{R_I}_{\rm eff}
$ \cite{pra062315ref38},
where $H^{R_0}_{\rm eff}=3\Omega_r^2(|111\rangle\langle111|+|T_{1p}\rangle\langle T_{1p}|)/\Delta+3\Omega_r^2(|000\rangle\langle000|$
$+|rrr\rangle\langle rrr|)/2\Delta+5\Omega_r^2(|110\rangle\langle110|+|101\rangle\langle101|+|011\rangle\langle011|+|pp0\rangle\langle pp0|+|p0p\rangle\langle p0p|+|0pp\rangle\langle 0pp|)/2\Delta+2\Omega_r^2(|100\rangle\langle100|+|010\rangle\langle010|+|001\rangle\langle001|)/\Delta$ stands for the Stark-shift term and $H^{R_I}_{\rm eff}=2\Omega_r^2(|110\rangle\langle pp0|+|101\rangle\langle p0p|+|011\rangle\langle0pp|)/\Delta
+2\sqrt{3}\Omega_r^2|111\rangle\langle T_{1p}|/\Delta+3\Omega_r^3|000\rangle\langle rrr|/2\Delta^2+{\rm H.c.}$ denotes the interaction term. We can find that all steady states of Zeno pumping but $|W\rangle$ are no longer stable due to the Rydberg pumping. These states are all pumped to the excited states and further decay to ground states $|0\rangle$ and $|1\rangle$ by the spontaneous emission of Rydberg states, which has been described by $L_i^{1(2)}$ and $L_i^{3(4)}$.
The detailed processes of the effective transitions have been illustrated in Fig.~\ref{effzeno}(b). There are four groups of ground states, $ \{ |000\rangle  \},\{ |111\rangle  \},\{ |011\rangle,|101\rangle,|110\rangle  \} $ and $\{ |001\rangle,|010\rangle,|100\rangle  \}$, which can be pumped to the three groups of excited states, $\{|rrr\rangle\},\{ |T_{1p}\rangle  \}$ and $\{ |pp0\rangle,|p0p\rangle,|0pp\rangle \}$ through effective classical fields with Rabi frequencies $3\Omega_r^3/2\Delta^2,2\sqrt{3}\Omega_r^2/\Delta$ and $2\Omega_r^2/\Delta$, respectively. Moreover, the excited states will spontaneously emit to the ground states again. The total system of Rydberg pumping will repeat the processes of pumping and decaying until it is stabilized into the unique steady state $|W\rangle$ and our purpose achieves.

\begin{figure*}
\centering
\includegraphics[scale=0.36]{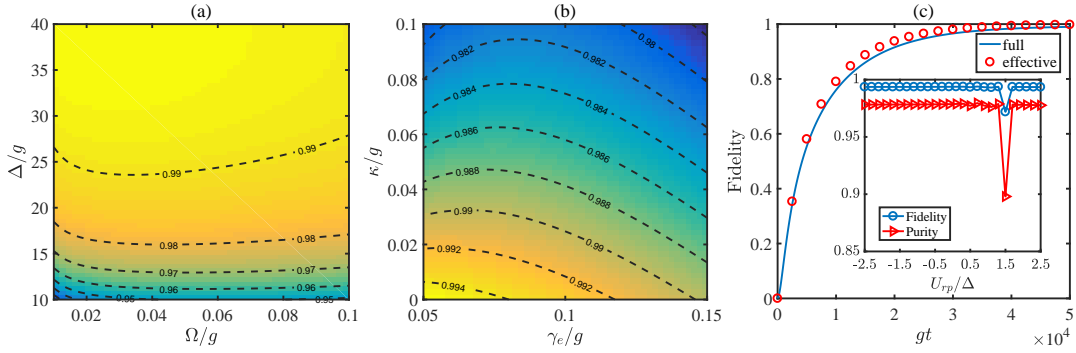}
\caption{\label{omegadelta}(a) Contour plot (dashed lines) of the steady-state fidelity of target state $|W\rangle$ as functions of $\Delta$ and $\Omega$. The relevant parameters are $\Omega_r=g,\gamma=0.002g$ and $\gamma_e=0.1g$.  (b) Contour plot (dashed lines) of the steady-state fidelity of $|W\rangle$ state with different $\kappa$ and $\gamma_e$. The parameter are $\Omega=0.01g,\Omega_r=g,\Delta=35g$ and $\gamma=\gamma_e/50$. (c) The time evolutions of fidelity for tripartite $W$ state governed by the full master equation (solid line) and the effective master equation (empty circles), where $\Omega=0.05g,\Omega_r=g,\Delta=45g,\gamma=0.002g,\gamma_e=0.1g$ and $\kappa=0$. The inset show the fidelity (empty circles) and the purity (empty triangles) with different $U_{rp}$. The definition of purity is $\mathcal{P}(t)={\rm Tr}[\rho_{ss}^2(t)]$. The parameters are $\Omega=0.05g,\Omega_r=g,\Delta=42g,\gamma=0.002g,\gamma_e=0.1g$ and $\kappa=0$.}
\end{figure*}
In Fig.~\ref{omegadelta}(a), we plot the steady-state fidelity of target state $|W\rangle$ as functions of $\Delta$ and $\Omega$ with a perfect cavity, where the fidelity is defined as $F=\sqrt{\langle W|\rho_{ss}(t)|W\rangle}$ and $\rho_{ss}(t)$ is the steady-state solution by solving the full master equation $\dot\rho=0$.  It forcefully proves the feasibility of the scheme that there are wide ranges of $\Delta$ and $\Omega$ to make the fidelity exceed $99\%$. Investigating the change of $\Omega/g$, we can find that, although the limit of Zeno pumping requires $g\gg\Omega$, it is not necessary that the less $\Omega/g$ the better the quality of target state.  In the range of $g\gg\Omega$, when the other parameters are fixed, there is a optimal value of $\Omega/g$ to maximise the fidelity of tripartite $W$ state. The scheme also need another limiting condition $\Delta\gg\Omega_r$ to achieve the Rydberg antiblockade. We have set $g=\Omega_r$ in the Fig.~\ref{omegadelta}(a). Hence the decreasing of $\Delta/g$ will destroy the limit of $\Delta\gg\Omega_r$ and lead to a decline in fidelity of tripartite $W$ state.

To discuss the robustness of the proposal against the loss of cavity and the influence of atomic spontaneous emission, a contour plot (dashed lines) of the steady-state fidelity of $|W\rangle$ state is shown with different $\kappa$ and $\gamma_e$ in Fig.~\ref{omegadelta}(b). The atomic spontaneous emission is an important resource in our protocol. Consequently, the  small $\gamma_e$ will reduce the fidelity of $|W\rangle$. On the other hand, since the quantum Zeno dynamics makes the cavity keep at the vacuum state, the fidelity can be still above $98\%$ even if $\kappa=0.1g$ and $0.058g<\gamma_e<0.12g$. It reflects the protocol is robust against the loss of cavity and also demonstrates the feasibility of the scheme.

In the following, we consider the dependence of the fidelity on $gt$ in Fig.~\ref{omegadelta}(c). The system is initialized at the state $|000\rangle|0\rangle_c$. The solid line represents the evolution of the target-state fidelity governed by the full master equation, which can reach $99\%$ at $gt=5\times10^4$. The fidelity of target state governed by the effective master equation (empty circles) is in good agreement with the solid line, which means the validity of the effective system. Therefore, the behavior of the actual system can be forecasted by the effective system.

In the above investigation, we have considered the Rydberg-mediated interactions satisfy $U_{rr}\gg U_{rp}$ with suitable principle quantum number and ignore the terms of $U_{rp}$. Even for large $U_{rp}$, we find the terms of $U_{rp}$ have no effect on the mission of Rydberg pumping and the state $|W\rangle$ is still the unique steady state of total system as long as $U_{rp}\neq1.5\Delta$. While $U_{rp}=1.5\Delta$, the transitions $|100\rangle\leftrightarrow|prr\rangle,|010\rangle\leftrightarrow|rpr\rangle$ and $|001\rangle\leftrightarrow|rrp\rangle$ will occur and break the principle of the Rydberg pumping, which can be written as $5\Omega_r^3(|100\rangle\langle prr|+|010\rangle\langle rpr|+|001\rangle\langle rrp|)/2\Delta^2+{\rm H.c.}$ and cause the $|W\rangle$ is no longer the steady state of the whole system. Then we estimate the character of system for different $U_{rp}$ by steady-state purity (empty triangles) in the inset of Fig.~\ref{omegadelta}(c).  When the system is in a pure steady state, the purity will be qual to unit. On the contrary, the purity will be less than unit. In the inset, the purity is around $97.95\%$ with all $U_{rp}$ except $U_{rp}=1.5\Delta$, which means the status of system is nearly a pure state except for the situation of $U_{rp}=1.5\Delta$. In order to further certify the above analysis about $U_{rp}$, we plot the steady-state fidelity of state $|W\rangle$ as a function of $U_{rp}$ (empty circles) in the inset. Similarly, the fidelity is almost constant with $99.4\%$ as long as $U_{rp}\neq1.5\Delta$. Combining the two curves of the inset, we successfully prove the validity of the above analysis and the general applicability of the present scheme.

Finally, we investigate the experimental feasibility. In experiment, the Rabi laser frequency $\Omega_r$ can be adjusted continuously between $2\pi\times(0,100)$ MHz \cite{prl090402,pra062315}. In \cite{brennecke2007}, Brennecke \textit{et al.} achieved the strong coupling of a Bose-Einstein condensate to the quantized filed of an ultrahigh-finesse optical cavity. The relevant cavity quantum electrodynamics (cavity QED) parameters were chosen as $(g,\kappa,\gamma_e)=2\pi\times(10.6,1.3,3.0)$ MHz. In \cite{klmoeref43}, the decay rate of Rydberg state was $\gamma=2\pi\times0.03$ MHz. Thus, we calculate the steady-state fidelity via these parameters, which can reach $98.42\%$ with $\Omega_r=2g,\Omega=0.002g$ and $\Delta=100g$. When we select another group of cavity QED parameters, $(g,\kappa,\gamma_e)=2\pi\times(185,53,3)$ MHz, provided by Volz \textit{et al.} in \cite{pra062315ref43}, and choose the decay rate of $20$D Rydberg state as $\gamma=2\pi\times0.144$ MHz \cite{pra052504}, the fidelity can be realized at $98.87\%$ with the other parameters, $\Omega_r=100$ MHz, $\Omega=0.002g $ and $\Delta=24g$. Furthermore, the fidelity can be enhanced to $99.09\%$ with the experimental parameters \cite{pra062315ref41} $(g,\kappa,\gamma_e,\gamma)=2\pi\times(14.4,0.66,3,0.03)~ {\rm MHz},\Omega_r=1.6g,\Omega=0.006g$ and $\Delta=80g$. In conclusion, the results plenarily exhibit the experimental feasibility of our scheme.

In summary, we successfully propose a dissipative scheme to generalize the tripartite $W$ state. The scheme can be equivalently separated into two simultaneous processes, which are Zeno pumping and Rydberg pumping, respectively. The Zeno pumping combines the quantum Zeno dynamics with atomic spontaneous emission to stabilize the system at mixed states of $|111\rangle,|000\rangle,|W'\rangle$ and $|W\rangle$. The Rydberg pumping utilizes the atomic spontaneous emission and antiblockade to break up the stability of the above states but $W$ state. { Combining the two pumps, the tripartite $W$ state becomes the unique state of the whole system, which means the target state can be achieved from an arbitrary initial state. Ultimately, the fidelity of target state can be above $98\%$ with the current experimental parameters. In experimental, the coupling of five trapped ions with cavity has been realized \cite{PhysRevLett.116.223001} and we believe our scheme supplies a new prospect to prepare multipartite  entanglements.}

X. Q. Shao would like to express
his thanks to Dr F. Reiter for his valuable
discussion. This work is supported by National Natural Science Foundation of China (NSFC) under Grants No. 11775048,
No. 11674049, No. 11774047.

\bibliographystyle{apsrev4-1}
\bibliography{Wr}

\end{document}